\documentclass{iopart}
\usepackage{iopams}

\newtheorem{Proposition}{Proposition}
\newtheorem{Lemma}{Lemma}
\newtheorem{Corollary}{Corollary}
\newtheorem{Definition}{Definition}

\newenvironment{proof}{\noindent{\bf proof.}}{\hspace{\stretch{1}}\opensquare}

\begin{document}

\title{Subspace preserving completely positive maps}
\author{Johan \AA berg}
\address{Department of Quantum Chemistry, Uppsala University,
 Box 518, SE-751 20 Uppsala, Sweden}
\ead{johan.aaberg@kvac.uu.se}

\begin{abstract}
A class of quantum channels and completely positive maps (CPMs) are
introduced and investigated. These, which we call \emph{subspace
preserving} (SP) CPMs has, in the case of trace preserving CPMs, a
simple interpretation as those which preserve probability weights on a
given orthogonal sum decomposition of the Hilbert space of a quantum
system. Several equivalent characterizations of SP CPMs are proved and
an explicit construction of all SP CPMs, is provided. For a subclass
of the SP channels a construction in terms of joint unitary evolution
with an ancilla system, is presented.
\end{abstract}


\section{Introduction}
Completely positive maps (CPMs) and trace preserving completely
positive maps \cite{Kraus} have important use as models of operations
on quantum systems. In this article a special type of CPMs, the
\emph{subspace preserving} CPMs, are defined and investigated. To some
extent the material presented here should be regarded as a toolbox, to
be used in future investigations, like in \cite{ref2} where
\emph{subspace local} trace preserving CPMs are introduced, and in
\cite{ref3} where the concept of \emph{gluings} of CPMs is
developed. Nevertheless, trace preserving SP CPMs do have a simple
conceptual interpretation.

Imagine some kind of box with impenetrable walls. It is assumed that
if a particle is put into such a box, it stays there; it neither
`leaks out' from the box, nor is it annihilated. Suppose we have two
such boxes and one \emph{single} particle. This particle can be put in
an arbitrary state in this box-pair. It may be localized in one of the
boxes, in a superposition, or any mixture of localized or delocalized
states. The question is: given the restriction that the boxes are
impenetrable to the particle, what kind of operations can we, in
principle, perform on the state of this particle? Put differently, if
the only restriction on the evolution is that there should be no
transfer of the particle between the boxes, what kind of evolution is
allowed, else allowing any type of interaction with environment or
between the boxes?  We search for the family of trace preserving CPMs
which obey the restriction of no `particle transfer' between the two
boxes.

The Hilbert space of the two-box system can be decomposed into an
orthogonal sum of two subspaces. One of these subspaces represents the
set of pure states localized in one of the boxes, the other subspace
represents the pure states localized in the other box. If $P_{1}$ is
the projector onto the subspace of localized pure states of box $1$,
and if $\Phi$ is the trace preserving CPM of the two-box system, then
the condition that the particle stays in box $1$ when put there, can
be formulated as $\Tr(P_{1}\Phi(\rho)) = \Tr(P_{1}\rho)$, where $\rho$
denotes the initial density operator. This can be interpreted as
conservation of probability; it is the same probability to find the
particle in box $1$, after the operation has been performed, as it was
before. This definition, or rather a wider definition which includes
more general types of situations, is used to derive some equivalent
characterizations of these types of CPMs and also to derive an
explicit expression for all such CPMs. In the above example, the SP
channels can be characterized as those which preserve certain 2-valued
observables, which is related to
\cite{Ozawa}.

The proofs presented here are all made under the limiting assumption
that all involved Hilbert spaces are finite-dimensional. This
assumption is made primarily to avoid mathematical
technicalities. Much of the material is likely to have analogies in
case of separable \cite{Krey} Hilbert spaces, with some technical
modifications. This is not treated here however.

The structure of this article is the following. In section \ref{se3}
the concept of subspace preserving CPMs is introduced and some
equivalent characterizations of this class of CPMs are proved. In
section \ref{se4} a special type of matrix representation of CPMs is
described. In section \ref{se5} the matrix representation of the
previous section is applied to SP CPMs. Expressions which makes it
possible generate all SP CPMs, are deduced. In section \ref{se6} we
turn to the special case of SP CPMs with identical source and target
spaces, and moreover identical decompositions of the source and target
spaces, to show a unitary representation for these CPMs. In section
\ref{sum} a summary is presented.
\section{Subspace preserving CPMs}
\label{se3}
We begin by establish some notation, terminology, and basic concepts
used throughout this article.  $\mathcal{H}$ denotes a
finite-dimensional complex Hilbert space. $\mathcal{H}$ with various
subscripts denotes the same.  The set of linear operators on
$\mathcal{H}$ is denoted $\mathcal{L}(\mathcal{H})$. Moreover,
$\mathcal{L}(\mathcal{H}_{S},\mathcal{H}_{T})$ denotes the set of
linear operators from $\mathcal{H}_{S}$ to $\mathcal{H}_{T}$. For two
Hermitian operators $A,B\in\mathcal{L}(\mathcal{H})$ we let $B\geq A$
denote $\langle\psi|B-A|\psi\rangle\geq 0$ for all
$|\psi\rangle\in\mathcal{H}$.

Given a linear map
$\phi:\mathcal{L}(\mathcal{H}_{S})\rightarrow\mathcal{L}(\mathcal{H}_{T})$,
we say that $\mathcal{H}_{S}$ is the \emph{source-space} of $\phi$ and
that $\mathcal{H}_{T}$ is the \emph{target space} of $\phi$ (or just
\emph{source} and \emph{target} for short). The source and target
space should not be confused with the domain and the range of
$\phi$. The domain of $\phi$ is $\mathcal{L}(\mathcal{H}_{S})$ and the
range is a subspace of $\mathcal{L}(\mathcal{H}_{T})$.  In this
investigation we are concerned with special linear maps $\phi$, the
\emph{completely positive maps} (CPM) \cite{Kraus}.  It has been shown
\cite{Kraus} that if the source and target space of a linear map
$\phi$ are separable, then $\phi$ is a CPM if and only if there exists
a sequence (finite or countable) of operators
$\{V_{k}\}_{k}\subset\mathcal{L}(\mathcal{H}_{S},\mathcal{H}_{T})$,
such that $\phi(Q) = \sum_{k}V_{k}Q V_{k}^{\dagger}$ for all
$Q\in\mathcal{L}(\mathcal{H}_{S})$ and fulfilling the condition
$\sum_{k}V_{k}^{\dagger}V_{k}\leq a\hat{1} $, where $0\leq
a<+\infty$. We say that $\{V_{k}\}_{k}$ is a \emph{Kraus
representation} of $\phi$. The Kraus representation $\{V_{k}\}_{k}$
potentially contains an infinite number of elements (also in the
finite-dimensional case) so that we strictly speaking have to define
what type of convergence we are considering in the sum
$\sum_{k}V_{k}QV_{k}^{\dagger}$. However, since the involved spaces
are assumed to be finite-dimensional, this is not such an involved
question. Moreover, if the target and sources are finite dimensional,
it is always possible to find finite Kraus representations (as will be
seen).

A CPM is called \emph{trace preserving} if $\Tr(\Phi(Q))=\Tr(Q)$ for
every trace class operator $Q$. Since the present analysis is
restricted to finite-dimensional Hilbert spaces, the set of trace
class operators coincide with the set of linear operators. To
emphasize that a CPM is trace preserving, we denote it with a Greek
capital letter, while Greek small letters denote general CPMs. The
word `channel' is here used as synonymous with trace preserving CPM.

In the following, when discussing CPMs, $\mathcal{H}_{S}$ denotes the
source space and $\mathcal{H}_{T}$ the target space of the CPM in
question, unless otherwise stated. These spaces are assumed to be
finite-dimensional. Moreover, the source and target spaces are
decomposed into orthogonal sums of subspaces as:
\begin{equation}
\label{st1st2}
\mathcal{H}_{S} = \mathcal{H}_{s1}\oplus \mathcal{H}_{s2},\quad 
\mathcal{H}_{T} = \mathcal{H}_{t1}\oplus \mathcal{H}_{t2},
\end{equation}
where $\mathcal{H}_{s1}$, $\mathcal{H}_{s2}$, $\mathcal{H}_{t1}$, and
$\mathcal{H}_{t2}$ are assumed to be at least
one-dimensional. Furthermore $P_{s1}$ denotes the projection operator
onto $\mathcal{H}_{s1}$, $P_{s2}$ the projection operator onto
$\mathcal{H}_{s2}$, and similarly for $P_{t1}$ and $P_{t2}$.

Now we are in position to define subspace preserving CPMs. The word
`preserving' refers to preservation of probability weight on selected
subspaces. Strictly speaking this terminology is a misnomer for CPMs
which are not trace preserving.
\begin{Definition}
\rm
Let $\phi$ be a CPM with source space $\mathcal{H}_{S}$ and target
space $\mathcal{H}_{T}$.  If $\phi$ fulfills both the conditions
\begin{equation}
\label{co1}
\fl\Tr(P_{t1}\phi(|\psi\rangle\langle\psi|)) = 0,
\quad\forall |\psi\rangle\in\mathcal{H}_{s2},\quad 
\Tr(P_{t2}\phi(|\psi\rangle\langle\psi|)) = 0,
\quad\forall |\psi\rangle\in\mathcal{H}_{s1},
\end{equation}
then $\phi$ is \emph{subspace preserving} (SP) from
$(\mathcal{H}_{s1},\mathcal{H}_{s2})$ to
$(\mathcal{H}_{t1},\mathcal{H}_{t2})$.
\end{Definition}
In proposition \ref{ekviv} it is shown that in case of trace
preserving CPMs, this definition is equivalent to a characterization
in line with the discussion in the introduction.

In this definition ``$|\psi\rangle\langle\psi|$'' and ``$\forall
|\psi\rangle\in\mathcal{H}_{sj}$'' can be replaced with ``$Q$'' and
``$Q\in\mathcal{L}(\mathcal{H}_{sj})$'', or with ``$\rho$'' and ``for
all density operators $\rho$ on $\mathcal{H}_{sj}$''. This because of
linearity of $\phi$ and the fact that any element $Q$ in
$\mathcal{L}(\mathcal{H}_{sj})$ can be written as a (complex) linear
combination of four density operators on $\mathcal{H}_{sj}$, which in
turn can be written as a sum of outer products of elements in
$\mathcal{H}_{sj}$.

\begin{Lemma}
\label{lenoll}
Let $\phi$ be a CPM.  If $\Tr(P_{ti}\phi(|\psi\rangle\langle\psi|)) =
0$ for all $|\psi\rangle\in\mathcal{H}_{sj}$ and if $\{V_{k}\}_{k}$ is
any Kraus representation of $\phi$, then $P_{ti}V_{k}P_{sj} =
0,\quad\forall k$.
\end{Lemma}
\begin{proof}
Let $|\psi\rangle\in\mathcal{H}_{sj}$ be arbitrary. Let
$|\chi\rangle\in\mathcal{H}_{ti}$ be an arbitrary normalized
vector. By $|\chi\rangle\langle\chi|\leq P_{ti}$ and by the positivity
of $\phi(|\psi\rangle\langle\psi|)$ follows
\begin{equation}
\label{leeq}
0 \leq \langle\chi|\phi(|\psi\rangle\langle\psi|)|\chi\rangle \leq
\Tr(P_{ti}\phi(|\psi\rangle\langle\psi|)) = 0.
\end{equation}
Let $\{V_{k}\}_{k}$ be an arbitrary Kraus representation of
$\phi$. From (\ref{leeq}) one obtains
$\sum_{k}|\langle\chi|V_{k}|\psi\rangle|^{2} = 0$, from which it
follows that $\langle\chi|V_{k}|\psi\rangle = 0,\quad\forall k$. Since
this is true for arbitrary $|\psi\rangle\in\mathcal{H}_{sj}$ and
arbitrary normalized $|\chi\rangle\in\mathcal{H}_{ti}$ it follows that
$P_{ti}V_{k}P_{sj}=0$.
\end{proof}
\begin{Proposition}
\label{sepkr}
Let $\phi$ be a CPM. If $\phi$ is SP from
$(\mathcal{H}_{s1},\mathcal{H}_{s2})$ to
$(\mathcal{H}_{t1},\mathcal{H}_{t2})$ then for every Kraus
representation $\{V_{k}\}_{k}$ of $\phi$, there exists operators
$V_{1,k}$ and $V_{2,k}$, such that
\begin{equation}
\label{decompoKr}
V_{k} = V_{1,k} + V_{2,k},\quad P_{t1}V_{1,k}P_{s1} = V_{1,k},\quad
P_{t2}V_{2,k}P_{s2} = V_{2,k},\quad\forall k.
\end{equation} 
Conversely if $\phi$ has a Kraus representation on the form
(\ref{decompoKr}), then $\phi$ is SP from
$(\mathcal{H}_{s1},\mathcal{H}_{s2})$ to
$(\mathcal{H}_{t1},\mathcal{H}_{t2})$.
\end{Proposition}
\begin{proof}
Let $\{V_{k}\}_{k}$ be any Kraus representation of $\phi$. If $\phi$
is assumed to be SP $(\mathcal{H}_{s1},\mathcal{H}_{s2})$ to
$(\mathcal{H}_{t1},\mathcal{H}_{t2})$, then by combining lemma
\ref{lenoll} with the two conditions (\ref{co1}) it is found that
$P_{t2}V_{k}P_{s1} = 0$ and $P_{t1}V_{k}P_{s2} = 0$. Define $V_{1,k}$
and $V_{2,k}$ by $V_{1,k}=P_{t1}V_{k}P_{s1}$ and
$V_{2,k}=P_{t2}V_{k}P_{s2}$. These fulfill the conditions stated in the
proposition.

The second statement follows since $\phi$, written on Kraus representation 
with  $\{V_{1,k}+V_{2,k}\}_{k}$, fulfills the conditions  (\ref{co1}). 
\end{proof}

An \emph{arbitrary} CPM $\phi$ with source $\mathcal{H}_{S}$ and
target $\mathcal{H}_{T}$, can be written
\begin{equation}
\label{allm}
\phi(Q) = \sum_{i=1,2}\sum_{j=1,2}\sum_{k=1,2}
\sum_{l=1,2}P_{ti}\phi(P_{sk}QP_{sl})P_{tj},
\quad\forall Q\in\mathcal{L}(\mathcal{H}_{S}).
\end{equation} 
This should be compared with the fourth statement of the following
proposition, which gives an analogous expression in case $\phi$ is SP
from $(\mathcal{H}_{s1},\mathcal{H}_{s2})$ to
$(\mathcal{H}_{t1},\mathcal{H}_{t2})$.
\begin{Proposition}
\label{transmit}
Let $\phi$ be a CPM with source $\mathcal{H}_{S}$ and target
$\mathcal{H}_{T}$. The following are equivalent
\begin{enumerate}
\item\label{trl1} $\phi$ is SP from 
$(\mathcal{H}_{s1},\mathcal{H}_{s2})$ to  
$(\mathcal{H}_{t1},\mathcal{H}_{t2})$.
\item\label{trl2} $P_{ti}\phi(Q)P_{tj} = \phi(P_{si}QP_{sj})
\quad i =1,2\quad j=1,2,\quad\forall Q\in \mathcal{L}(\mathcal{H}_{S})$.
\item\label{trl3} $P_{ti}\phi(P_{sk}QP_{sl})P_{tj} = 
\delta_{ik}\delta_{lj}P_{ti}\phi(P_{si}QP_{sj})P_{tj}
\quad i,j,k,l = 1,2,\quad\forall Q\in \mathcal{L}(\mathcal{H}_{S})$.
\item\label{trl4} $\phi(Q) = 
\sum_{i=1,2}\sum_{j=1,2}P_{ti}\phi(P_{si}QP_{sj})P_{tj},
\quad\forall Q\in\mathcal{L}(\mathcal{H}_{S})$. 
\end{enumerate}
\end{Proposition}
\begin{proof}
  (\ref{trl1}) $\Rightarrow$ (\ref{trl2}): From the existence of a
  Kraus representation $\{V_{1,k}+V_{2,k}\}_{k}$ given by proposition
  \ref{sepkr}, follows (\ref{trl2}) , since
  $P_{ti}\phi(Q)P_{tj}=\sum_{k}V_{i,k}QV_{j,k}^{\dagger}=\phi(P_{si}QP_{sj})$.

(\ref{trl2}) $\Rightarrow$ (\ref{trl3}): By (\ref{trl2}) follows
$P_{ti}\phi(P_{sk}QP_{sl})P_{tj}=\phi( P_{si}P_{sk}QP_{sl}P_{sj})$,
from which (\ref{trl3}) follows.

(\ref{trl3}) $\Rightarrow$ (\ref{trl4}): By inserting the condition
(\ref{trl3}) into equation (\ref{allm}), (\ref{trl4}) follows.

(\ref{trl4}) $\Rightarrow$ (\ref{trl1}): Assuming $\phi$ fulfills the
condition of (\ref{trl4}) then $\phi$ fulfills conditions (\ref{co1})
and hence, $\phi$ is SP from $(\mathcal{H}_{s1},\mathcal{H}_{s2})$ to
$(\mathcal{H}_{t1},\mathcal{H}_{t2})$.
\end{proof}
\begin{Proposition}
\label{compoSP}
If a CPM $\phi_{a}$ is SP from $(\mathcal{H}_{s1},\mathcal{H}_{s2})$
to $(\mathcal{H}_{t1},\mathcal{H}_{t2})$ and if a CPM $\phi_{b}$ is SP
from $(\mathcal{H}_{t1},\mathcal{H}_{t2})$ to
$(\mathcal{H}_{r1},\mathcal{H}_{r2})$ then $\phi_{b}\circ\phi_{a}$ is
SP from $(\mathcal{H}_{s1},\mathcal{H}_{s2})$ to
$(\mathcal{H}_{r1},\mathcal{H}_{r2})$.
\end{Proposition}
The spaces $\mathcal{H}_{r1}$ and $\mathcal{H}_{r2}$ are assumed to be
finite-dimensional, at least one-dimensional and being orthogonal
complements of each other.

\begin{proof}
Using proposition \ref{transmit} follows
\begin{equation}
\fl P_{ri}\phi_{b}\circ\phi_{a}(Q)P_{rj} = 
\phi_{b}(P_{ti}\phi_{a}(Q)P_{tj}) =\phi_{b}(\phi_{a}(P_{si}QP_{sj})),
\quad\forall Q\in\mathcal{L}(\mathcal{H}_{S}). 
\end{equation}
Hence, by proposition \ref{transmit}, $\phi_{b}\circ\phi_{a}$ is SP.
\end{proof}
\begin{Proposition}
\label{ekviv}
Let $\Phi$ be a trace preserving CPM. The following are equivalent
\begin{enumerate}
\item \label{ekl1} $\Phi$ is SP from 
$(\mathcal{H}_{s1},\mathcal{H}_{s2})$ to $(\mathcal{H}_{t1},\mathcal{H}_{t2})$.
\item \label{ekl2}
$\Tr(P_{t1}\phi(Q)) = \Tr(P_{s1}Q) ,\quad\forall Q\in
\mathcal{L}(\mathcal{H}_{S})$.
\item \label{ekl3}
$\Tr(P_{t2}\phi(Q)) = \Tr(P_{s2}Q) ,\quad\forall Q\in
\mathcal{L}(\mathcal{H}_{S})$.
\end{enumerate}
\end{Proposition}
\begin{proof}
(\ref{ekl2}) $\Leftrightarrow$ (\ref{ekl3}): Using the fact that
$P_{s1}+P_{s2} = \hat{1}_{S}$ and similarly for the target space, one
can write
\begin{displaymath}
\Tr(\Phi(Q)) - \Tr(Q) = \Tr(P_{t1}\Phi(Q)) - \Tr(P_{s1}Q) + 
 \Tr(P_{t2}\Phi(Q)) - \Tr(P_{s2}Q).
\end{displaymath} 
Since $\Phi$ is trace preserving, it follows that $\Tr(P_{t1}\Phi(Q))
- \Tr(P_{s1}Q)= -\Tr(P_{t2}\Phi(Q)) + \Tr(P_{s2}Q)$, from which the
equivalence of (\ref{ekl2}) and (\ref{ekl3}) is obtained.

(\ref{ekl2}) $\Rightarrow$ (\ref{ekl1}): With $Q =
|\psi\rangle\langle\psi|$, $|\psi\rangle\in\mathcal{H}_{s2}$ into
(\ref{ekl2}), the first of the conditions (\ref{co1}) is seen to
hold. We know (\ref{ekl2}) $\Rightarrow$ (\ref{ekl3}). Hence, with
$|\psi\rangle\in\mathcal{H}_{s1}$ into (\ref{ekl3}), the second of the
conditions (\ref{co1}) is seen to hold. Hence, $\Phi$ is SP.

(\ref{ekl1}) $\Rightarrow$ (\ref{ekl2}): From proposition
\ref{transmit} it follows that
$P_{t1}\Phi(Q)P_{t1}=\Phi(P_{s1}QP_{s1})$. Hence, $\Tr(P_{t1}\Phi(Q))=
\Tr(P_{t1}\Phi(Q)P_{t1})= \Tr(\Phi(P_{s1}QP_{s1}))= \Tr(P_{s1}Q)$,
where the last equality follows from $\Phi$ being trace preserving.
\end{proof}
\section{Matrix representation of CPMs}
\label{se4}
In this section some material is presented which will be useful in the
analysis of SP and SL CPMs, as well as for the gluing concept. We here
discuss a type of matrix representation of CPMs, where CPMs are
represented by positive semi-definite matrices. One can construct many
different types of matrix representations, and the choice is a matter
of convenience, depending on the application. The type used here has
appeared before in the literature (for examples see
\cite{ChNi,Lid}). Another form can be found in \cite{Choi}. In order
to establish the properties of the representation used here, we state
and prove the following proposition.
\begin{Proposition}
\label{rep1}
Let
$\{V_{m}\}_{m=1}^{M}\subset\mathcal{L}(\mathcal{H}_{S},\mathcal{H}_{T})$
with $M=\dim(\mathcal{H}_{S})\dim(\mathcal{H}_{T})$ be a basis of
$\mathcal{L}(\mathcal{H}_{S},\mathcal{H}_{T})$. Let the elements of
the set $\{\phi_{m,m'}\}_{m,m'=1}^{M}$ be defined as
\begin{equation}
\phi_{mm'}(Q) = V_{m}QV_{m'}^{\dagger},
\quad\forall Q \in\mathcal{L}(\mathcal{H}_{S}).
\end{equation}
The set $\{\phi_{mm'}\}_{m,m'=1}^{M}$ forms a basis of
$\mathcal{L}(\mathcal{L}(\mathcal{H}_{S}),\mathcal{L}(\mathcal{H}_{T}))$
and hence the equation \begin{equation}
\label{Fmphi} 
\phi(Q) = \sum_{m,m'=1}^{M}F_{m,m'}V_{m}Q V_{m'}^{\dagger},
\quad\forall Q \in\mathcal{L}(\mathcal{H}_{S}),
\end{equation}
 defines a linear bijection between the set of complex $M\times M$
 matrices $F =[F_{m,m'}]_{m,m'}$ and
 $\mathcal{L}(\mathcal{L}(\mathcal{H}_{S}),\mathcal{L}(\mathcal{H}_{T}))$.

Moreover, equation (\ref{Fmphi}) defines a bijection between the set
of all positive semi-definite matrices $F$ and the set of all CPMs
with source space $\mathcal{H}_{S}$ and target space
$\mathcal{H}_{T}$.
\end{Proposition}
In this type of representation we regard $\phi$ as a vector in
$\mathcal{L}(\mathcal{L}(\mathcal{H}_{S}),\mathcal{L}(\mathcal{H}_{T}))$,
and represent it via a basis in this space. This basis we construct
using an arbitrary basis of $\mathcal{L}(\mathcal{H}_{S},\mathcal{H}_{T})$.
 This representation
may be compared with more `classical' matrix representations of linear
maps. In such a representation we would use a basis of
$\mathcal{L}(\mathcal{H}_{S})$ and a basis of
$\mathcal{L}(\mathcal{H}_{T})$ to represent linear maps from
$\mathcal{L}(\mathcal{H}_{S})$ to $\mathcal{L}(\mathcal{H}_{T})$ as
matrices.

\begin{proof}
Most of the proposition follows immediately if it can be shown that
$\{\phi_{mm'}\}_{mm'}$ is a basis of
$\mathcal{L}(\mathcal{L}(\mathcal{H}_{S}),\mathcal{L}(\mathcal{H}_{T}))$.
Clearly each $\phi_{mm'}$ belongs to this set. The set
$\{\phi_{mm'}\}_{mm'}$ contains $M^{2}$ elements, hence it suffices to
show that it is linearly independent (since
$\dim(\mathcal{L}(\mathcal{L}(\mathcal{H}_{S}),\mathcal{L}(\mathcal{H}_{T})))
= M^{2}$.) Define $\alpha= \sum_{mm'}C_{mm'}\phi_{mm'}$, for some
arbitrary complex numbers $C_{mm'}$. It has to be shown that
$\alpha=0$ implies $C_{mm'}=0$ for all $m,m'$. Let
$|\psi\rangle,|\chi\rangle\in\mathcal{H}_{S}$,
$|\eta\rangle\in\mathcal{H}_{T}$ all be arbitrary. The condition
$\alpha=0$ implies $\alpha(|\psi\rangle\langle\chi|)|\eta\rangle = 0$.
This can be rewritten (by inserting the definition of $\phi_{mm'}$ and
rearrange) as
$\sum_{m}(\sum_{m'}C_{mm'}\langle\chi|V_{m'}^{\dagger}|\eta\rangle)
V_{m}|\psi\rangle = 0$. Since $|\psi\rangle$ is arbitrary, it follows
that
$\sum_{m}(\sum_{m'}C_{mm'}\langle\chi|V_{m'}^{\dagger}|\eta\rangle)V_{m}=0$.
Since $\{V_{m}\}_{m}$ is a linearly independent set, the last equation
implies $\sum_{m'}C_{mm'}\langle\chi|V_{m'}^{\dagger}|\eta\rangle =
0$, for each $m$. Since $|\chi\rangle$ and $|\eta\rangle$ are
arbitrary, it further follows that $\sum_{m'}C_{mm'}V_{m'}^{\dagger} =
0,\quad\forall m$. If $\{V_{m}\}_{m}$ is a linearly independent set,
then so is $\{V_{m}^{\dagger}\}_{m}$ and hence,
$C_{mm'}=0,\quad\forall m,m'$. Hence, $\{\phi_{mm'}\}_{mm'}$ is a
basis of
$\mathcal{L}(\mathcal{L}(\mathcal{H}_{S}),\mathcal{L}(\mathcal{H}_{T}))$.
 From $\{\phi_{mm'}\}_{mm'}$ being a basis it follows that the complex
numbers $F_{mm'}$ in (\ref{Fmphi}) are the expansion coefficients of
$\phi$ with respect to this basis. Hence, the bijectivity stated in
the proposition follows.

It remains to show the bijectivity between the set of positive
semi-definite matrices F and the set of CPMs. Assuming $F$ is positive
semi-definite, there exists some unitary matrix $U$, such that
$U^{\dagger}DU = F$, where $D$ is a diagonal matrix $D
=[d_{m}\delta_{m,m'}]_{m,m'}$ with $d_{m}\geq 0$. Define
$\{W_{n}\}_{n}$ by $W_{n}=\sqrt{d_{n}}\sum_{m}U^{*}_{n,m}V_{m}$ for
$n$ for which $d_{n}\neq 0$. (For this proof $d_{n}\neq 0$ is not
needed, but will be useful in a later proof.) One can check that
$\{W_{n}\}_{n}$ so defined is a Kraus representation of $\phi$. Since
any element in
$\mathcal{L}(\mathcal{L}(\mathcal{H}_{S}),\mathcal{L}(\mathcal{H}_{T}))$
which has a Kraus representation, is a CPM \cite{Kraus}, it follows
that $\phi$ is a CPM.

It remains to show that for any CPM $\phi$ in
$\mathcal{L}(\mathcal{L}(\mathcal{H}_{S}),\mathcal{L}(\mathcal{H}_{T}))$,
the corresponding matrix $F$ is positive semi-definite.  Let
$\{|s_{n}\rangle\}_{n=1}^{N}$ be an arbitrary orthonormal basis of
$\mathcal{H}_{S}$. Let $|\psi\rangle=
\sum_{n}|s_{n}\rangle|s_{n}\rangle$. (Hence, $|\psi\rangle$ is an
element of $\mathcal{H}_{S}\otimes\mathcal{H}_{S}$.) Let $I_{N}$
denote the identity CPM with source and target $\mathcal{H}_{S}$.
Since $\phi$ is a CPM it follows, by definition \cite{Kraus}, that
$\phi\otimes I_{N}$ maps positive semi-definite operators to positive
semi-definite operators. Let
$A\in\mathcal{L}(\mathcal{H}_{S},\mathcal{H}_{T})$ be
arbitrary. Clearly $(A^{\dagger}\otimes \hat{1}_{S})[\phi\otimes
I_{N}](Q)(A\otimes \hat{1}_{S})$ is a positive semi-definite operator
for any positive semi-definite $Q$. One may verify that
$\langle\psi|(A^{\dagger}\otimes \hat{1}_{S})[\phi\otimes
I_{N}](|\psi\rangle\langle\psi|)(A\otimes \hat{1}_{S})|\psi\rangle =
\sum_{mm'}F_{mm'}\Tr(A^{\dagger}V_{m})\Tr(V_{m'}^{\dagger}A)$ and
hence
\begin{equation}
\label{samb}
\sum_{mm'}F_{mm'}\Tr(A^{\dagger}V_{m})\Tr(V_{m'}^{\dagger}A)\geq 0,
\quad\forall A\in\mathcal{L}(\mathcal{H}_{S},\mathcal{H}_{T}).
\end{equation}
 Since $\{V_{m}\}_{m=1}^{M}$ is a linearly independent set, it follows
 that the matrix $F$ has to be positive semi-definite. To see this,
 one can use that $(A,B)=\Tr(A^{\dagger}B)$ is an inner product (the
 Hilbert-Schmidt inner product \cite{HS}) on
 $\mathcal{L}(\mathcal{H}_{S},\mathcal{H}_{T})$. With respect to this
 inner product we may form a Gram-matrix \cite{LanTis} with elements
 $G_{mm'}= \Tr(V_{m}^{\dagger}V_{m'})$. Since $\{V_{m}\}_{m=1}^{M}$ is
 a linearly independent set, this matrix is positive definite
 \cite{LanTis} and hence invertible. Hence, if $c =
 (c_{m})_{m=1}^{M}\in\mathbb{C}$ is arbitrary, we may construct $A=
 \sum_{m=1}^{M}c_{m}V_{m}$. By inserting this into equation
 (\ref{samb}) one finds that $c^{\dagger}GFG^{\dagger}c\geq 0$ for any
 $c\in\mathbb{C}^{M}$. Hence, $GFG^{\dagger}\geq 0$. Since $G$ is
 invertible it follows that $F\geq 0$, which shows the proposition.
\end{proof}
\begin{Proposition}
\label{corolnum}
To every CPM $\phi$ there exists a linearly independent Kraus
representation. The number of elements $K(\phi)$ in a linearly
independent Kraus representation only depends on the CPM and not on
the choice of linearly independent Kraus representation. The number
$K(\phi)$, to be called the Kraus number, have the following
properties:
\begin{enumerate}
\item $K(\phi)\leq\dim(\mathcal{H}_{S})\dim(\mathcal{H}_{T})$.
\item $K(\phi)$ is the minimal number of operators needed in any Kraus 
representation of $\phi$. 
\item $K(\phi)$ is the number of non-zero eigenvalues, 
counted with multiplicity of the matrix $F$ given by proposition \ref{rep1}, 
for any choice of basis of  $\mathcal{L}(\mathcal{H}_{S},\mathcal{H}_{T})$.
\end{enumerate} 
\end{Proposition} 
\begin{proof}
First it has to be proved that for every CPM it is possible to find a
\emph{linearly independent} Kraus representation. Actually we have
already constructed such a set in the proof of proposition
\ref{rep1}. The set $\{W_{n}\}_{n}$, defined in that proof, is such a
set. To see why this is the case one can note how $\{W_{n}\}_{n}$ was
constructed. By assumption $\{V_{m}\}_{m=1}^{M}$ forms a basis. Hence,
the set $\{\widetilde{W}_{n}\}_{n=1}^{M}$ defined by
$\widetilde{W}_{n} = \sum_{m=1}^{M}U^{*}_{n,m}V_{m}$ for the unitary
matrix $U$, must also be a basis and hence linearly independent. Since
the elements $W_{n}$ were defined as $W_{n} =
\sqrt{d_{n}}\widetilde{W}_{n}$, for the \emph{non-zero} $d_{n}$, it
follows that $\{W_{n}\}_{n}$ must also be a linearly independent
set. Hence, there exists a linearly independent Kraus representation.

The number of elements in $\{W_{n}\}_{m}$ is equal to the number of
non-zero eigenvalues $d_{n}$ (counted with multiplicity) of the matrix
$F$. Moreover, there cannot be more than
$\dim(\mathcal{H}_{S})\dim(\mathcal{H}_{T})$ non-zero eigenvalues.

Consider the same CPM $\phi$, but represented with respect to some
other choice of basis $\{\widetilde{V}_{m}\}_{m=1}^{M}$ of
$\mathcal{L}(\mathcal{H}_{S},\mathcal{H}_{T})$. As with any change of
basis, the new basis and the original are related via an invertible
matrix $A$ as $V_{m} = \sum_{m'}\widetilde{V}_{m'}A_{m'm}$. When
inserting this into (\ref{Fmphi}), one finds that the new matrix
$\widetilde{F}$ is related to the old as $\widetilde{F} =
AFA^{\dagger}$. Since $A$ is invertible, the number of non-zero
eigenvalues of $F$ and $\widetilde{F}$ are the same. The eigenvalues
per se may change, but not the number of non-zero eigenvalues. This
follows from ``Sylvester's law of inertia'' \cite{LanTis}, since $F$
and $\widetilde{F}$ are \emph{congruent} ($\widetilde{F} =
AFA^{\dagger}$ for some non-singular $A$) and Hermitian
\cite{LanTis}. Hence, the number of non-zero eigenvalues of the matrix
$F$ is independent of the choice of basis. From this follows directly
that the number of operators in a linearly independent Kraus
representation is independent of the choice of linearly independent
representation. Hence, it is possible to define $K(\phi)$ as the
number of operators in a linearly independent Kraus representation. It
also follows that if a Kraus representation has $K(\phi)$ elements, it
has to be a linearly independent Kraus representation. Moreover, there
cannot be any Kraus representation with less than $K(\phi)$ elements.
\end{proof}

One may wonder about the nature of the set of all linearly independent
Kraus representations of a CPM.
\begin{Proposition}
\label{allli}
Let $\phi$ be a CPM with Kraus number $K = K(\phi)$. Let
$\{V_{k}\}_{k=1}^{K}$ be an arbitrary but fixed linearly independent
Kraus representation of $\phi$. Let $\{V'_{k}\}_{k=1}^{K}$ be defined
by
\begin{equation}
\label{vdef}
V'_{k} =\sum_{k'=1}^{K}U_{kk'}V_{k'},\quad k=1,\ldots,K.
\end{equation}
Equation (\ref{vdef}) defines a bijection between the set of all
linearly independent Kraus representations of $\phi$ and the set of
unitary $K\times K$ matrices $U=[U_{kk'}]_{k,k'=1}^{K}$.
\end{Proposition}
Note that in this proposition, two linearly independent Kraus
representations $\{V_{k}\}_{k=1}^{K}$ and $\{V'_{k}\}_{k=1}^{K}$ are
regarded as `equal', if and only if $V_{k}=V'_{k}$ for all
$k=1,\ldots,K$. Hence, two linearly independent Kraus representations
differing only in the ordering of its elements are regarded as
`different' in this proposition.

\begin{proof}
 It is a well known result \cite{Preskill}, \cite{NilChu} that any two
 Kraus representations of the same CPM can be connected via a unitary
 matrix (where a possibly smaller set of Kraus operators is padded
 with zero operators, such that the two sets have the same number of
 elements), and moreover that any unitary matrix $U$ creates a new
 Kraus representation. By combining these facts with proposition
 \ref{corolnum} it follows that every unitary $K\times K$ matrix $U$
 gives a linearly independent Kraus representation via (\ref{vdef})
 and moreover that every linearly independent Kraus representation can
 be reached in this manner. Hence, (\ref{vdef}) is surjective from the
 set of unitary $K\times K$ matrices $U$ to the set of linearly
 independent Kraus representations. Hence, it remains to show that the
 mapping is injective. Suppose $U$ and $U'$ both give the same
 linearly independent Kraus representation. Then
 $\sum_{k=1}^{K}U_{k',k}V_{k}=\sum_{k=1}^{K}U'_{k',k}V_{k}$ for all
 $k'=1,\ldots,K$. Since $\{V_{k}\}_{k=1}^{K}$ is a linearly
 independent set, it follows that $U_{k',k} = U'_{k',k}$ for all
 $k,k'$. Hence, no two distinct unitary matrices are mapped to the
 same linearly independent Kraus representation. Hence, the mapping is
 injective.
\end{proof}

We conclude this section by proving that it is always possible to find
a linearly independent Kraus representation based on an operator set
which is orthonormal with respect to the Hilbert-Schmidt inner product
$(A,B) = \Tr(A^{\dagger}B)$ \cite{HS}.
\begin{Proposition}
Let $\phi$ be a CPM with Kraus number $K=K(\phi)$. There exists a
linearly independent Kraus representation on the form
$\{\sqrt{r_{n}}Y_{n}\}_{n=1}^{K}$ where each $r_{n}$ is a positive
real number $r_{n}>0$ and where the set $\{Y_{n}\}_{n=1}^{K}$ is
orthonormal with respect to the Hilbert-Schmidt inner product.
\end{Proposition}
\begin{proof}
Let $\{V_{m}\}_{m}$ be a linearly independent Kraus representation of
$\phi$.  Form the Gram-Matrix $R = [R_{m,m'}]_{m,m'}$, with elements
$R_{m,m'}= \Tr(V_{m}^{\dagger}V_{m'})$. Since $\{V_{m}\}_{m=1}^{M}$ is
a linearly independent set, the Gram matrix is positive definite
\cite{LanTis}. Hence, there exists a unitary $M\times M$ matrix $U$
and numbers $r_{n}>0$, such that
$\sum_{m,m'}U_{m,n}^{*}R_{m,m'}U_{m'n'} = r_{n}\delta_{n,n'}$. Let
$\{\widetilde{Y}_{n}\}_{n=1}^{N}$ be the linearly independent Kraus
representation defined by $\widetilde{Y}_{n}
=\sum_{m}V_{m}U_{m,n}$. By construction
$\Tr(\widetilde{Y}_{n}^{\dagger}\widetilde{Y}_{n'})
=r_{n}\delta_{nn'}$. Since $r_{n}>0$ one may define $Y_{n} =
\frac{1}{\sqrt{r_{n}}}\widetilde{Y}_{n}$. The set
$\{\sqrt{r}_{n}Y_{n}\}_{n=1}^{N}$ so defined, fulfills the statements
of the proposition.
\end{proof}
\section{Matrix representation of SP CPMs}
\label{se5}
Here the matrix representation presented in the previous section is
applied to the case of SP CPMs.
\begin{Proposition}
\label{issrep}
Let $\{V_{k}\}_{k}$ be a basis of
$\mathcal{L}(\mathcal{H}_{s1},\mathcal{H}_{t1})$ and let
$\{W_{l}\}_{l}$ be a basis of
$\mathcal{L}(\mathcal{H}_{s2},\mathcal{H}_{t2})$. A CPM $\phi$ is SP
from $(\mathcal{H}_{s1},\mathcal{H}_{s2})$ to
$(\mathcal{H}_{t1},\mathcal{H}_{t2})$ if and only if it can be written
\begin{eqnarray}
\label{matrep}
 \phi(Q) = & \sum_{k,k'}F_{k,k'}V_{k}QV_{k'}^{\dagger}
 +\sum_{k,l'}F_{k,l'}V_{k}QW_{l'}^{\dagger} \nonumber\\ &+
 \sum_{l,k'}F_{l,k'}W_{l}QV_{k'}^{\dagger} +
 \sum_{l,l'}F_{l,l'}W_{l}QW_{l'}^{\dagger},\quad\forall
 Q\in\mathcal{L}(\mathcal{H}_{S}),
\end{eqnarray}
where the matrix $F$ is defined by
\begin{equation}
\label{submatrix}
F = \left[\begin{array}{cc}
\left[F_{k,k'}\right]_{kk'} & \left[F_{k,l'}\right]_{kl'}\\
\left[F_{l,k'}\right]_{lk'} & \left[F_{l,l'}\right]_{ll'}
\end{array}\right],
\end{equation}
is positive semi-definite.  Moreover, (\ref{matrep}) defines a
bijection between the set of such positive semi-definite matrices $F$
and the set of CPMs that are SP from
$(\mathcal{H}_{s1},\mathcal{H}_{s2})$ to
$(\mathcal{H}_{t1},\mathcal{H}_{t2})$.
\end{Proposition}
\begin{proof}
We begin with the ``if'' part of the proposition.  The set
$\{V_{k}\}_{k}\cup\{W_{l}\}_{l}$ is \emph{not} a basis of
$\mathcal{L}(\mathcal{H}_{S},\mathcal{H}_{T})$. In order to use
proposition \ref{rep1} we have to complete it with a basis of
$\mathcal{L}(\mathcal{H}_{s1},\mathcal{H}_{t2})$:
$\{Y_{n}^{12}\}_{n}$, and a basis of
$\mathcal{L}(\mathcal{H}_{s2},\mathcal{H}_{t1})$:
$\{Y_{m}^{21}\}_{m}$. The set $B =
\{V_{k}\}_{k}\cup\{W_q{l}\}_{l}\cup\{Y_{n}^{12}\}_{n}\cup\{Y_{m}^{21}\}_{m}$
forms a basis of $\mathcal{L}(\mathcal{H}_{S},\mathcal{H}_{T})$. The
matrix (\ref{submatrix}) is a sub-matrix of the matrix $F$ given by
proposition \ref{rep1}, with respect to the basis $B$. Moreover,
(\ref{submatrix}) forms the only non-zero part of the matrix. Hence,
the matrix $F$ of proposition \ref{rep1} is positive semi-definite if
and only if the sub-matrix (\ref{submatrix}) is. Hence, $\phi$ defined
by (\ref{matrep}), is a CPM if and only if (\ref{submatrix}) is
positive semi-definite. By inserting $\phi$ as defined by
(\ref{matrep}), into the conditions (\ref{co1}), it follows that
$\phi$ is an SP CPM.

For the ``only if'' part, assume $\phi$ is any CPM which is SP from
$(\mathcal{H}_{s1},\mathcal{H}_{s2})$ to
$(\mathcal{H}_{t1},\mathcal{H}_{t2})$.  Let $F$ be the matrix (given
by proposition \ref{rep1}) which represents $\phi$ with respect to the
basis $B$: $\phi(Q) = \sum_{ss'}F_{ss'}X_{s}QX_{s'}^{\dagger}$, where
each $X_{s}$ is an element of type $V$, $W$, $Y^{12}$ or
$Y^{21}$. Using proposition \ref{transmit} one can show that
$F_{ss'}=0$ if $X_{s}$ or $X_{s'}$ is an element of type $Y^{12}$ or
$Y^{21}$. Hence, if $\phi$ is an SP CPM, then the only (potentially)
non-zero part of the sum $\sum_{ss'}F_{ss'}X_{s}QX_{s'}^{\dagger}$ is
(\ref{matrep}), where the sub-matrix (\ref{submatrix}) consists of
those elements $F_{ss'}$ for which $X_{s}$ and $X_{s'}$ are both of
the type $V$ or $W$. Hence, we have shown that (\ref{matrep}) defines
a surjective map from the set of positive semi-definite matrices
(\ref{submatrix}) to the set of SP CPMs. That this map is also
injective follows from the bijectivity stated in proposition
\ref{rep1}.
\end{proof}
\begin{Corollary}
If $\phi$ is SP CPM from $(\mathcal{H}_{s1},\mathcal{H}_{s2})$ to
$(\mathcal{H}_{t1},\mathcal{H}_{t2})$ then
\begin{displaymath}
K(\phi)\leq \dim(\mathcal{H}_{s1})\dim(\mathcal{H}_{t1}) +
\dim(\mathcal{H}_{s2})\dim(\mathcal{H}_{t2}).
\end{displaymath}
\end{Corollary}
\begin{proof}
The corollary follows directly from the characterization of the Kraus
number $K(\phi)$ as the number of non-zero eigenvalues of the
representation matrix $F$ of proposition \ref{rep1}. By proposition
\ref{issrep} only the sub-matrix (\ref{submatrix}) is non-zero. Since
this sub-matrix is a $[\dim(\mathcal{H}_{s1})\dim(\mathcal{H}_{t1}) +
\dim(\mathcal{H}_{s2})\dim(\mathcal{H}_{t2})]
\times[\dim(\mathcal{H}_{s1})\dim(\mathcal{H}_{t1})
+ \dim(\mathcal{H}_{s2})\dim(\mathcal{H}_{t2})]$ matrix, the corollary
follows.
\end{proof}

For the rest of this section we develop means to rewrite proposition
\ref{issrep}. This is done by characterizing positive semi-definite
matrices in terms of sub-matrices.
\begin{Lemma}
\label{geninek}
If $D$ is a complex positive semi-definite $N\times N$ matrix, then
\begin{displaymath}
|a^{\dagger}D b|^{2}\leq a^{\dagger}Da\,\, b^{\dagger}Db ,\quad\forall
a,b\in\mathbb{C}^{N}.
\end{displaymath} 
\end{Lemma}
Note that $a$ and $b$ are regarded as being column vectors.

\begin{proof}
Consider the $2\times 2$ matrix $F$
\begin{displaymath}
F =\left[\begin{array}{cc}
a^{\dagger}Da  & a^{\dagger}Db \\
b^{\dagger}Da & b^{\dagger}Db
\end{array}\right].
\end{displaymath}
Let $c\in\mathbb{C}^{2}$ be arbitrary. ($c$ is regarded as a column
vector with elements $c_{1}$ and $c_{2}$.)  $c^{\dagger}Fc =
(c_{1}^{*}a^{\dagger}+c_{2}^{*}b^{\dagger})D(c_{1}a+c_{2}b)\geq 0$,
where the last inequality follows since $D$ is positive
semi-definite. Since $c$ is arbitrary it follows that $F$ is positive
semi-definite. By the fact that the determinant of a positive
semi-definite matrix is non-negative, the statement of the lemma
follows, since $\det(F)=a^{\dagger}Da\,\, b^{\dagger}Db-|a^{\dagger}D
b|^{2}$.
\end{proof}
\begin{Lemma}
\label{condposdef}
Let $A$ and $B$ be $N\times N$ and $M\times M$ positive semi-definite
matrices, respectively. Let $C$ be a complex $N\times M$ matrix and
let $F$ be the $(N+M)\times (N+M)$ matrix
\begin{displaymath}
F = \left[\begin{array}{cc}
A & C\\
C^{\dagger} & B
\end{array}\right].
\end{displaymath}
Then $F$ is positive semi-definite if and only if 
\begin{equation}
\label{cond2}
P_{A,0}C = 0,\quad CP_{B,0} = 0,\quad A\geq CB^{\ominus}C^{\dagger},
\end{equation}
where $B^{\ominus}$ denotes the Moore-Penrose pseudo inverse of
$B$. $P_{A,0}$ denotes the orthogonal projector onto the zero
eigenspace of $A$ and analogously for $P_{B,0}$.  In (\ref{cond2}),
the condition $\quad A\geq CB^{\ominus}C^{\dagger}$ can be replaced
with the condition $B \geq C^{\dagger}A^{\ominus}C$.
\end{Lemma}
A comment on the Moore-Penrose pseudo inverse (MP-inverse) \cite{MPi},
\cite{MPi2}, \cite{LanTis} is perhaps suitable here. If a matrix is
invertible, its MP-inverse reduces to the ordinary inverse. If a
(finite) square matrix $B$ is Hermitian, and if its \emph{non-zero}
eigenvalues are $\lambda_{k}$, with corresponding orthonormal
eigenvectors $b_{k}$, such that $B =
\sum_{k}\lambda_{k}b_{k}b_{k}^{\dagger}$ (regard $b_{k}$ as column
vectors) then $B^{\ominus} =
\sum_{k}\lambda_{k}^{-1}b_{k}b_{k}^{\dagger}$. The operator
$BB^{\ominus}=B^{\ominus}B$ is the projector onto the range of $B$.
Moreover, $P_{A,0} = I_{N}-AA^{\ominus}$ and $P_{B,0} =
I_{M}-BB^{\ominus}$.  In the following the MP-inverse will be used
without further comments.

\begin{proof}
Throughout this proof, elements of $\mathbb{C}^{N}$, $\mathbb{C}^{M}$,
$\mathbb{C}^{N+M}$ are all regarded as column vectors.  We begin by
proving that if $F$ is positive semi-definite then $P_{A,0}C = 0$,
$CP_{B,0} = 0$ and $A \geq CB^{\ominus}C^{\dagger}$. For any $q_{A}$
in the zero eigenspace of $A$, it follows by lemma \ref{geninek}
applied to the matrix $F$, that $q_{A}^{\dagger}Cq_{B}=0$ for every
$q_{B}\in\mathbb{C}^{M}$. From this follows $P_{A,0}C = 0$. Similarly
one can derive $CP_{B,0} = 0$. Let $q\in \mathbb{C}^{N+M}$ be
arbitrary and let $q_{A}$ denote the projection of $q$ onto the first
$N$ components of $q$ and let $q_{B}$ denote the projection of $q$
onto the rest of the $M$ components ($q = q_{A}\oplus q_{B}$). Then
\begin{equation}
\label{dec}
q^{\dagger}Fq = q^{\dagger}_{A}Aq_{A} + q^{\dagger}_{A}Cq_{B}
+q^{\dagger}_{B}C^{\dagger}q_{A} + q^{\dagger}_{B}Bq_{B}.
\end{equation}
Let $q$ be such that $q_{B} = -B^{\ominus}C^{\dagger}q_{A}$ and
$q_{A}\in \mathbb{C}^{N}$ being arbitrary. By inserting this $q$ into
(\ref{dec}) one finds $q^{\dagger}Fq =
q^{\dagger}_{A}(A-CB^{\ominus}C^{\dagger})q_{A}$. Since $F$ is assumed
to be positive semi-definite, it follows that
$q^{\dagger}_{A}(A-CB^{\ominus}C^{\dagger})q_{A}\geq 0$. Since $q_{A}$
is arbitrary one obtains $A-CB^{\ominus}C^{\dagger}\geq 0$. By an
analogous reasoning it can be shown that $B \geq
C^{\dagger}A^{\ominus}C$. (Let $q_{B}\in\mathbb{C}^{M}$ be arbitrary
and $q_{A} = -A^{\ominus}Cq_{B}$).

Next we have to prove that if $P_{A,0}C = 0$, $CP_{B,0} = 0$ and $A
\geq CB^{\ominus}C^{\dagger}$, then $F$ is positive semi-definite. Let
$q\in\mathbb{C}^{N+M}$ be arbitrary and let $q = q_{A}\oplus q_{B}$.
From $A \geq CB^{\ominus}C^{\dagger}$ follows that
$q_{A}^{\dagger}Aq_{A} \geq
q_{A}^{\dagger}CB^{\ominus}C^{\dagger}q_{A}$.  By inserting the last
expression into equation (\ref{dec}) one obtains
\begin{eqnarray}
\label{le13}
 q^{\dagger}Fq &\geq & q_{A}^{\dagger}CB^{\ominus}C^{\dagger}q_{A} +
 q^{\dagger}_{A}Cq_{B} +q^{\dagger}_{B}C^{\dagger}q_{A} +
 q^{\dagger}_{B}Bq_{B}\nonumber\\ &= &
 q_{A}^{\dagger}C\sqrt{B}^{\ominus}\sqrt{B}^{\ominus}C^{\dagger}q_{A}
 + q^{\dagger}_{A}C\sqrt{B}^{\ominus}\sqrt{B}q_{B}\nonumber\\ & &+
 q^{\dagger}_{B}\sqrt{B}\sqrt{B}^{\ominus}C^{\dagger}q_{A} +
 q^{\dagger}_{B} \sqrt{B} \sqrt{B}q_{B}.
\end{eqnarray}
In the last equality above, it has been used that
$\sqrt{B}^{\ominus}\sqrt{B} = \sqrt{B}\sqrt{B}^{\ominus}$ is the
projector onto the range of $B$. Hence, $C
=C\sqrt{B}^{\ominus}\sqrt{B}$, since $CP_{B,0} = 0$.  Let $q_{D}=
\sqrt{B}^{\ominus}C^{\dagger}q_{A}$ and $q_{E} = \sqrt{B}q_{B}$. By
inserting the last expression into (\ref{le13}), one obtains
$q^{\dagger}Fq \geq q_{D}^{\dagger}q_{D} + q_{D}^{\dagger}q_{E} +
q_{E}^{\dagger}q_{D} + q_{E}^{\dagger}q_{E} = ||q_{D}+q_{E}||^{2}\geq
0$. Since $q$ is arbitrary, it follows that $F$ is positive
semi-definite.  A similar derivation can be done for $B \geq
C^{\dagger}A^{\ominus}C$.
\end{proof}

Now we are in position to state the reformulation of proposition
\ref{issrep} by which we end this section. One may wonder why this
reformulation is useful. Loosely speaking, proposition \ref{issrep} is
to prefer when we `search' the whole set of SP CPMs (with respect to
some choice of source and target decomposition) without any additional
assumptions. Proposition \ref{alliss} is more useful when considering,
not the whole set of SP CPMs, but subsets for which the sub-matrices
$A$ and $B$ are fixed. Proposition \ref{alliss} is particularly useful
if $A$ and $B$ can be chosen to have a simple form, like for example
identity matrices. (Indeed such conditions, or similar, do occur
\cite{ref3}.) Hence, depending on the specific problem at hand, one
of propositions \ref{issrep} or \ref{alliss} may be the preferable
tool.

By combination of proposition \ref{issrep} and lemma \ref{condposdef}
the following is obtained.
\begin{Proposition}
\label{alliss}
Let $\{V_{k}\}_{k=1}^{K}$ be a basis of
$\mathcal{L}(\mathcal{H}_{s1},\mathcal{H}_{t1})$, $K =
\dim\mathcal{H}_{s1}\dim\mathcal{H}_{t1}$, and let
$\{W_{l}\}_{l=1}^{L}$ be a basis of
$\mathcal{L}(\mathcal{H}_{s2},\mathcal{H}_{t2})$, $L =
\dim\mathcal{H}_{s2}\dim\mathcal{H}_{t2}$. Then $\phi$ defined by
\begin{equation}
\label{allagl2}
\fl \phi(Q) =  \sum_{kk'}A_{k,k'}V_{k}QV_{k'}^{\dagger}+
\sum_{ll'}B_{l,l'}W_{l}QW_{l'}^{\dagger} + 
\sum_{kl}C_{kl}V_{k}QW_{l}^{\dagger} + 
\sum_{kl}C_{kl}^{*}W_{l}QV_{k}^{\dagger},
\end{equation}
for all $Q\in\mathcal{L}(\mathcal{H}_{S})$, is an SP CPM from
$(\mathcal{H}_{s1},\mathcal{H}_{s2})$ to
$(\mathcal{H}_{t1},\mathcal{H}_{t2})$, if and only if the matrices $A
= [A_{k,k'}]_{k,k'}$, $B = [B_{l,l'}]_{l,l'}$ and $C =
[C_{k,l}]_{k,l}$ fulfill the relations
\begin{equation}
\label{kondi0}
A\geq 0,\quad B\geq 0
\end{equation}
\begin{equation}
\label{kondi}
P_{A,0}C = 0,\quad CP_{B,0} = 0,\quad A \geq CB^{\ominus}C^{\dagger},
\end{equation}
with $P_{A,0}$ and $P_{B,0}$ defined as in lemma \ref{condposdef}.

Moreover, (\ref{allagl2}) defines a bijection between the set of all
CPMs which are SP from $(\mathcal{H}_{s1},\mathcal{H}_{s2})$ to
$(\mathcal{H}_{t1},\mathcal{H}_{t2})$, and the set of all triples of
matrices $(A,B,C)$ fulfilling the conditions (\ref{kondi0}) and
(\ref{kondi}).
\end{Proposition}
\section{Unitary representation of a subclass of the SP channels}
\label{se6}
In this section a special case of SP channels is considered, the case
of identical source and target spaces and moreover with the orthogonal
decomposition of the target and source space being
equal. ($\mathcal{H}_{T} =\mathcal{H}_{S}$, $\mathcal{H}_{t1}
=\mathcal{H}_{s1}$, $\mathcal{H}_{t2} =\mathcal{H}_{s2}$.) The
`two-box system' described in the introduction is one example of such
a system.  To have a more comfortable terminology; if a CPM $\phi$ is
SP from $(\mathcal{H}_{s1},\mathcal{H}_{s2})$ to
$(\mathcal{H}_{s1},\mathcal{H}_{s2})$, then we say that $\phi$ is SP
\emph{on} $(\mathcal{H}_{s1},\mathcal{H}_{s2})$.

The representation presented here is of the form of a unitary
evolution of the system and an ancillary system. Given a channel with
identical source and target spaces there always exists a representation
in terms of a unitary evolution on the system and an ancilla system
\cite{Kraus}. As such, the representation presented here is nothing
new. The important aspect is rather the special \emph{form} of this
unitary representation. The meaning of this form of representation
perhaps become a bit more clear in \cite{ref2}. In \cite{ref2} a
unitary representation, for a class of channels called local subspace
preserving channels, is developed. These form a subset of the set of
SP channels.  One may compare the unitary representation for local
subspace preserving channels, with the representation presented in
proposition \ref{unitrepSP}. This comparison suggests that the local
subspace preserving channels consists precisely of the locally acting
SP channels, in the sense of \cite{ref2}. For more details the reader
is referred to \cite{ref2} and \cite{ref3}.
\begin{Proposition}
\label{unitrepSP}
Let $\Phi$ be a trace preserving CPM. $\Phi$ is SP on
$(\mathcal{H}_{s1},\mathcal{H}_{s2})$ if and only if there exists an
ancilla space $\mathcal{H}_{a}$, a normalized state
$|a\rangle\in\mathcal{H}_{a}$, and operators $V_{1}$ and $V_{2}$ on
$\mathcal{H}_{S}\otimes\mathcal{H}_{a}$, such that
\begin{equation}
\label{partis}
V_{1}V_{1}^{\dagger}  =  V_{1}^{\dagger}V_{1} = P_{s1}\otimes\hat{1}_{a},\quad
V_{2}V_{2}^{\dagger}  =  V_{2}^{\dagger}V_{2} = P_{s2}\otimes\hat{1}_{a},
\end{equation}
\begin{equation}
\label{unitrep}
\Phi(Q) = \Tr_{a}(U Q\otimes|a\rangle\langle a| U^{\dagger}),\quad
\forall Q\in \mathcal{L}(\mathcal{H}_{S}),
\end{equation}
where $U$ is the unitary operator $U = V_{1}+V_{2}$.
\end{Proposition}
\begin{proof}
First it has to be shown that $U$ is a unitary operator.
The conditions (\ref{partis}) imply
\begin{equation}
\label{subsp}
(P_{s1}\otimes\hat{1}_{a})V_{1}(P_{s1}\otimes\hat{1}_{a}) =
V_{1},\quad (P_{s2}\otimes\hat{1}_{a})V_{2}(P_{s2}\otimes\hat{1}_{a})
= V_{2}.
\end{equation}
This can be shown by using a singular value decomposition
\cite{LanTis} of $V_{1}$. $V_{1} =
\sum_{k}r_{k}|\psi_{k}\rangle\langle\eta_{k}|$, where
$\{|\psi_{k}\rangle\}_{k}$ and $\{|\eta_{k}\rangle\}_{k}$ both form
orthonormal sets of vectors, and $r_{k}>0$. By inserting this
decomposition into (\ref{partis}), it follows that
$|\psi_{k}\rangle,|\eta_{k}\rangle\in
\mathcal{H}_{s1}\otimes\mathcal{H}_{a}$. Hence,
$V_{1}$ fulfills (\ref{subsp}). An analogous reasoning holds for
$V_{2}$. Using (\ref{subsp}) and (\ref{partis}) the unitarity of $U$
follows.

To prove the ``if'' part we first note that $\Phi$ is trace preserving
by the form of equation (\ref{unitrep}). Moreover,
\begin{eqnarray}
\Tr(P_{s1}\Phi(Q)) & = 
\Tr\left((P_{s1}\otimes\hat{1}_{a})UQ\otimes|a\rangle\langle a|
U^{\dagger}\right)\nonumber\\
                  & = \Tr(V_{1} Q\otimes|a\rangle\langle a|
                  V_{1}^{\dagger}) = \Tr(P_{s1}Q).
\end{eqnarray}
According to proposition \ref{ekviv} this implies that $\Phi$ is SP on
$(\mathcal{H}_{s1},\mathcal{H}_{s2})$.

Now we turn to the proof of the ``only if'' part of the
proposition. By proposition \ref{sepkr} there exists a Kraus
representation of $\Phi$ on the form $\{V_{1,k}+V_{2,k}\}_{k}$ where
\begin{equation}
\label{bev1}
P_{s1}V_{1,k}P_{s1}=V_{1,k}, \quad P_{s2}V_{2,k}P_{s2}=V_{2,k}.
\end{equation}
Since $\Phi$ is trace preserving, it follows that
$\sum_{k}(V_{1,k}+V_{2,k})^{\dagger}(V_{1,k}+V_{2,k}) =
\sum_{k}V_{1,k}^{\dagger}V_{1,k}+\sum_{k}V_{2,k}^{\dagger}V_{2,k}=
\hat{1}$, which together with (\ref{bev1}) imply
\begin{equation}
\label{bev3}
\sum_{k}V_{1,k}^{\dagger}V_{1,k}= P_{s1}, \quad
 \sum_{k}V_{2,k}^{\dagger}V_{2,k}= P_{s1}.
\end{equation}
From proposition \ref{corolnum} it follows that
$\{V_{1,k}+V_{2,k}\}_{k}$ can be chosen such that it has finitely many
elements. Let this number be $K$.  Let
$\{|a_{0}\rangle\}\cup\{|a_{k}\rangle\}_{k=1}^{K}$ be an orthonormal
basis of an ancilla space $\mathcal{H}_{a}$. (If $K$ is the number of
Kraus operators, the ancilla space is of dimension $K+1$.) For $i=1,2$
define the following operators:
\begin{eqnarray}
\label{Vdef}
V_{i} = & P_{si}\otimes\hat{1}_{a} - P_{si}\otimes|a_{0}\rangle\langle
a_{0}| -
\sum_{kk'}V_{i,k}V_{i,k'}^{\dagger}\otimes|a_{k}\rangle\langle
a_{k'}|\nonumber\\ & + \sum_{k}V_{i,k}\otimes|a_{k}\rangle\langle
a_{0}|+ \sum_{k}V_{i,k}^{\dagger}\otimes|a_{0}\rangle\langle a_{k}|.
\end{eqnarray}
Using equations (\ref{bev3}) one may verify that $V_{1}$ and $V_{2}$
fulfill conditions (\ref{partis}). Moreover, one can check, using
$V_{1,k}V_{1,k'}^{\dagger}+V_{2,k}V_{2,k'}^{\dagger} =
(V_{1,k}+V_{2,k})(V_{1,k'}+V_{2,k'})^{\dagger}$, that
\begin{eqnarray}
\label{bev4}
U = & V_{1}+V_{2} = \hat{1}\otimes\hat{1}_{a} -
\hat{1}\otimes|a_{0}\rangle\langle a_{0}|\nonumber\\ & -
\sum_{kk'}(V_{1,k}+V_{2,k})(V_{1,k'}+V_{2,k'})^{\dagger}\otimes
|a_{k}\rangle\langle
a_{k'}|\nonumber\\ & +
\sum_{k}(V_{1,k}+V_{2,k})\otimes|a_{k}\rangle\langle a_{0}|+
\sum_{k}(V_{1,k}+V_{2,k})^{\dagger}\otimes|a_{0}\rangle\langle a_{k}|
\end{eqnarray}
and that
\begin{displaymath}
\Tr_{a}(U Q\otimes|a_{0}\rangle\langle a_{0}| U^{\dagger}) = 
\sum_{k}(V_{1,k}+V_{2,k})Q(V_{1,k}+V_{2,k})^{\dagger} = \Phi(Q),
\end{displaymath}
which proves the proposition.
\end{proof}
\section{Summary}
\label{sum}
A special family of completely positive maps (CPMs), named
\emph{subspace preserving} (SP), is introduced. In case of trace
preserving CPMs (channels) these can be characterized as those which
preserve probability weight on chosen subspaces. More specifically,
let $\Phi$ be a trace reserving CPM which maps density operators on
the finite-dimensional Hilbert space $\mathcal{H}_{S}$ to density
operators on the finite-dimensional Hilbert space
$\mathcal{H}_{T}$. (We say that $\mathcal{H}_{S}$ is the \emph{source
space} of $\Phi$, and $\mathcal{H}_{T}$ the \emph{target space} of
$\Phi$). Let $P_{s1}$ be the projection operator onto some, at least
one-dimensional, subspace $\mathcal{H}_{s1}$ of $\mathcal{H}_{S}$ and
let $P_{t1}$ be the projection operator onto some, at least
one-dimensional, subspace $\mathcal{H}_{t1}$ of $\mathcal{H}_{T}$. The
channel $\Phi$ is said to be subspace preserving from
$(\mathcal{H}_{s1},\mathcal{H}_{s2})$ to
$(\mathcal{H}_{t1},\mathcal{H}_{t2})$ if $\Tr(P_{t1}\Phi(\rho)) =
\Tr(P_{s1}\rho)$ for all density operators $\rho$ on
$\mathcal{H}_{S}$. In the actual analysis a more general definition is
used which covers also the case of not trace preserving CPMs.

Under the limiting assumption of finite-dimensional source and target
spaces, several equivalent characterizations of these types of
channels are deduced. Moreover, an expression which makes it possible
to generate all CPMs which are SP with respect to some orthogonal
decompositions of source and target spaces, is proved (proposition
\ref{issrep} and \ref{alliss}). In the special case of identical
source and target spaces $\mathcal{H}_{T}=\mathcal{H}_{S}$ and
moreover identical orthogonal decomposition of the source and target
$\mathcal{H}_{t1}=\mathcal{H}_{s1}$,
$\mathcal{H}_{t2}=\mathcal{H}_{s2}$, a representation in terms of a
special form of joint unitary evolution with an ancilla system, is
deduced. As a part of the analysis, the concepts of linearly
independent Kraus representations and the \emph{Kraus number} of CPMs
are introduced. This investigation is the first in a family of
articles investigating properties of channels with respect to
orthogonal decompositions of source and target spaces. The subsequent
members in this family is \cite{ref2} and \cite{ref3}, where some of
the material presented here is used.  
\ack I thank Erik Sj\"oqvist for
many valuable comments and discussions on the manuscript. I also thank
Marie Ericsson for discussions which started the train of thoughts
leading to this investigation. Finally I thank Osvaldo Goscinski for
reading and commenting the text.  

 \end{document}